\documentclass[british,english,letterpaper,twocolumn,superscriptaddress,aps]{revtex4}

\usepackage{graphicx,color,subfigure}
\usepackage{amssymb,amsmath}

\begin{document}

\title{Anisotropic viscoelastic phase separation in polydisperse hard rods: non-sticky gelation}

\author{Claudia Ferreiro-C\'{o}rdova}
\affiliation{School of Chemistry, Cantock's Close, University of Bristol, BS8 1TS, UK}
\affiliation{Centre for Nanoscience and Quantum Information, Tyndall Avenue, Bristol BS8 1FD, UK}
\affiliation{Laboratoire de Physique des Solides, Universit\'e Paris-Sud \& CNRS, UMR 8502, 91405 Orsay, France}

\author{C. Patrick Royall}
\affiliation{School of Chemistry, Cantock's Close, University of Bristol, BS8 1TS, UK}
\affiliation{Centre for Nanoscience and Quantum Information, Tyndall Avenue, Bristol BS8 1FD, UK}
\affiliation{H.H. Wills Physics Laboratory, Tyndall Ave., Bristol, BS8 1TL, UK}
\affiliation{Department of Chemical Engineering, Kyoto University, Kyoto 615-8510, Japan}

\author{Jeroen S. van Duijneveldt}
\affiliation{School of Chemistry, Cantock's Close, University of Bristol, BS8 1TS, UK}

\date{\today}

\begin{abstract}
Spinodal demixing of systems into two phases having very different viscosities leads to viscoelastic networks, i.e. gels. Here we consider demixing in a colloidal system where one phase is a nematic liquid crystal with a strongly anisotropic viscosity. We combine real space analysis of a sepiolite clay system with molecular dynamics computer simulation. We find a long-lived network with many of the characteristics of a colloidal gel. Remarkably the anisotropic viscosity means that flow is possible within the nematic network, but only along the network, which thus preserves its structure, leading to a new form of anisotropic gel. Our system presents a further novel property, in that it is comprised of hard particles. Thus our dynamically anisotropic gels form in the absence of attraction between the particles.  Thus we show that our new system exhibits the ingredients of gelation and conclude that it represent a new class of material, \emph{non-sticky} gels in which attractions are not present.
\end{abstract}

\maketitle

\section{Introduction}

Gelation, the emergence of a network of arrested material with finite zero-shear modulus upon slight cooling is among the most striking everyday features of condensed matter \cite{zaccarelli2007} and is an example of viscoelastic phase separation where a contrast in viscosity between the two phases is crucial to the formation of a long-lived network \cite{tanaka2000viscoelastic}. Gels can be soft (and biological) materials such as proteins \cite{cardinaux2007}, clays \cite{jabbarifarouji2007}, foods \cite{tanaka2013fara}, hydrogels \cite{helgeson2012} and tissues \cite{rose2014}. In addition a more diverse range of materials including granular matter \cite{ulrich2009}, phase-demixing oxides \cite{bouttes2014} and metallic glass formers \cite{baumer2013} also exhibit gelation. The mechanical properties of gels are influenced by their structure both locally \cite{hsiao2012,zhang2012} and at a macroscopic level through percolation of particles \cite{valadezperez2013} and network topology \cite{varrato2012}.

Despite its widespread occurrence, a complete understanding of gelation remains a challenge \cite{zaccarelli2007}. Two properties unify particulate gels so far produced. Firstly, the constituent particles experience an \emph{attraction} to one another (which may be effective, induced for example by depletion effects from added polymer \cite{poon2002} or DNA \cite{varrato2012}). Secondly, with some exceptions \cite{zhang2012}, both phases are \emph{isotropic}. The role played by the interparticle attraction is intimately related to the phase behaviour. In the case of gelation, spinodal decomposition (immediate demixing) leads to a phase rich in particles which is of sufficient colloid volume fraction
that its high viscosity results in a long-lived percolating network in which full demixing is suppressed  \cite{zaccarelli2007,tanaka2000viscoelastic,lu2008,royall2018}. Thus colloidal gels exhibit \emph{dynamical contrast} between the phases formed through spinodal decomposition \cite{zhang2013}.

Here we depart radically from this paradigm: we use a system of polydisperse hard rods which feature no meaningful attractions. Nevertheless in such a system, spinodal decomposition occurs from a thermodynamically unstable isotropic fluid to an isotropic fluid and nematic colloidal liquid crystal \cite{vanbruggen1999,ni2010,lettinga2006}.

But what of the dynamics -- why should the phase separation in our system arrest such that a gel forms? Crucially, the nematic phase exhibits an \emph{anisotropic viscosity}. Although the viscosity along the director is comparable to that in the isotropic phase, perpendicular to the director, the viscosity is around three orders of magnitude higher for our system, which is comparable to the (isotropic) dynamic contrast found in systems of attractive spheres \cite{zhang2013,royall2018}. We thus argue that polydisperse hard rods in which there is no attraction can in principle feature some of the properties required for spinodal gelation. Specifically, these amount to spinodal demixing (which produces a percolating network of the nematic phase). As for the dynamics,  we shall show that the rods align parallel to the ``arms'' of the gel, and thus while there may be flow along the arms, flow perpendicular to the arms is very strongly suppressed.

\begin{figure*}
\begin{center}
\centerline{\includegraphics[width=140 mm]{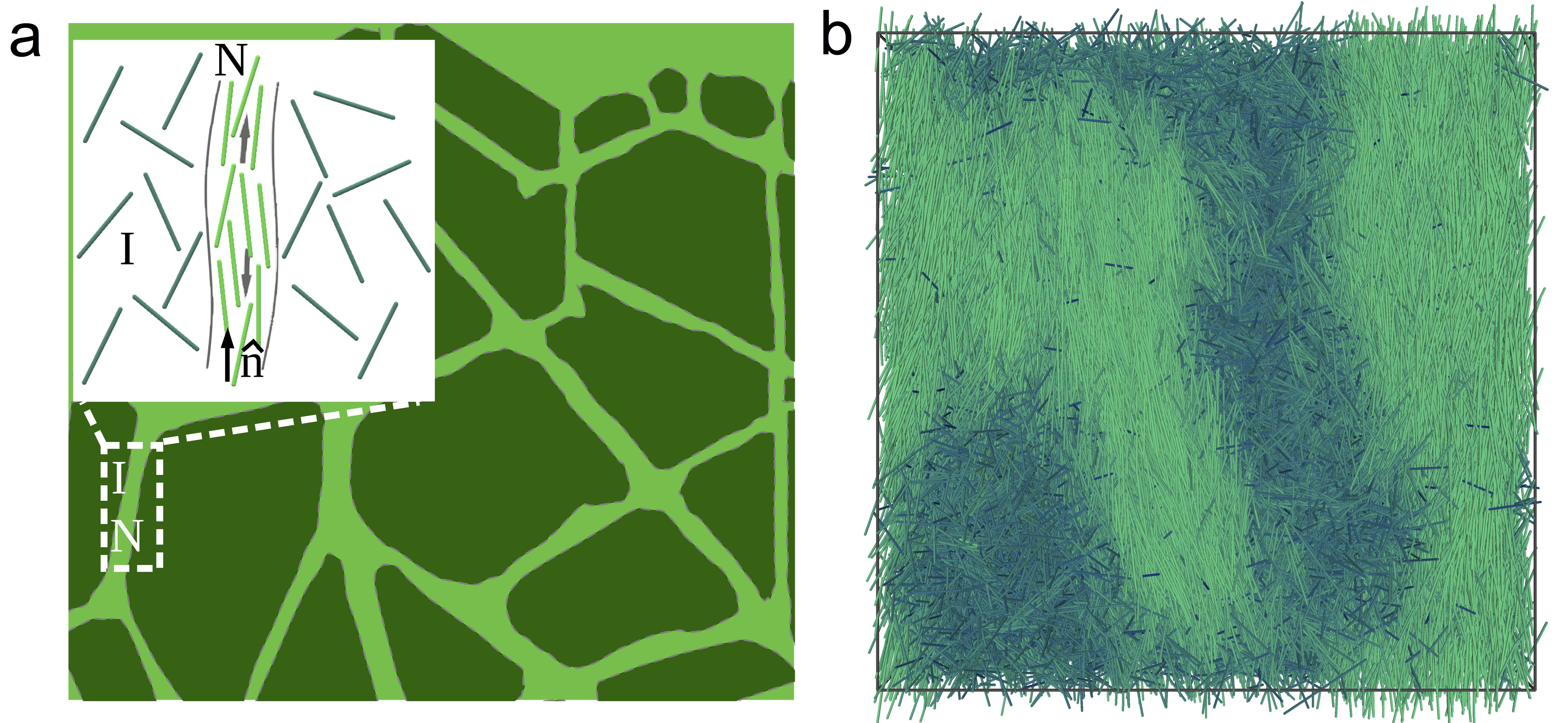}}
\caption{\label{figRodGelSchematic} 
\textbf{Schematic illustration of anisotropic viscoelastic phase separation.}
Spinodal decomposition leads to a percolating network of isotropic (I) and nematic (N) (which is bicontinuous in 3d). 
(a) Inset shows that rods align parallel to the local axis of the network. We expect that the viscosity along the local nematic axis is low enough to permit flow, while perpendicular to the director, flow is strongly suppressed. 
(b) Snapshot of a thin slice of a simulation box of rods at $\phi=0.140$. The colours indicate the local order parameter $S_i$ of each rod $i$ (see Appendix), and range from dark blue for isotropic rods to green for nematic rods.}
\end{center}
\end{figure*}

\section{Experimental and simulation}

Our colloidal rods have a mean aspect ratio $L/D=24.6$ and exhibit two features which are important here. Firstly they are hard, that is to say there is no evidence for any attraction \cite{woolston2015}. This means that any gelation behaviour we find occurs without attractions or even effective attractions such as those found in colloid-polymer mixtures \cite{poon2002}. Secondly, the rods are rather polydisperse in length $(35 \%  \rightarrow L=723 \pm 254$ nm) leading to a large gap in density between isotropic and nematic phases at phase coexistence. While this coexistence gap might shrink on long timescales due to segregation of the polydisperse rods \cite{speranza2002}, this is suppressed by the slow dynamics of the nematic phase and no evidence of segregation is seen on the experimental timescale.

For our simulations, we modelled the experimental systems using rods with no attraction between them. The dynamics analysis were carried out using 65000 rods with polydispersity only in length. Our model particles follow a Gaussian distribution with an average rod length of $L=24.48 \sigma \pm 9.06 \sigma$. Three different packing fractions are explored here for this system: $\phi =0.012$ (isotropic), $\phi=0.145$ (coexistence) and $\phi=0.357$ (nematic). For $\phi =0.145$ we found a coexistence phase with around $40\%$ of the rods in a strong nematic phase. The snapshots showed in the present work correspond to a bigger system of 180000 rods made with an average rod length of $L=24.48 \sigma \pm 9.08 \sigma$ and a packing fraction $\phi=0.140$ which corresponds to a coexistence phase. Further details, for both simulations and experiments, may be found in the Methods and Appendix.

\begin{figure}[htb]
\begin{center}
\centerline{\includegraphics[width=0.5\textwidth]{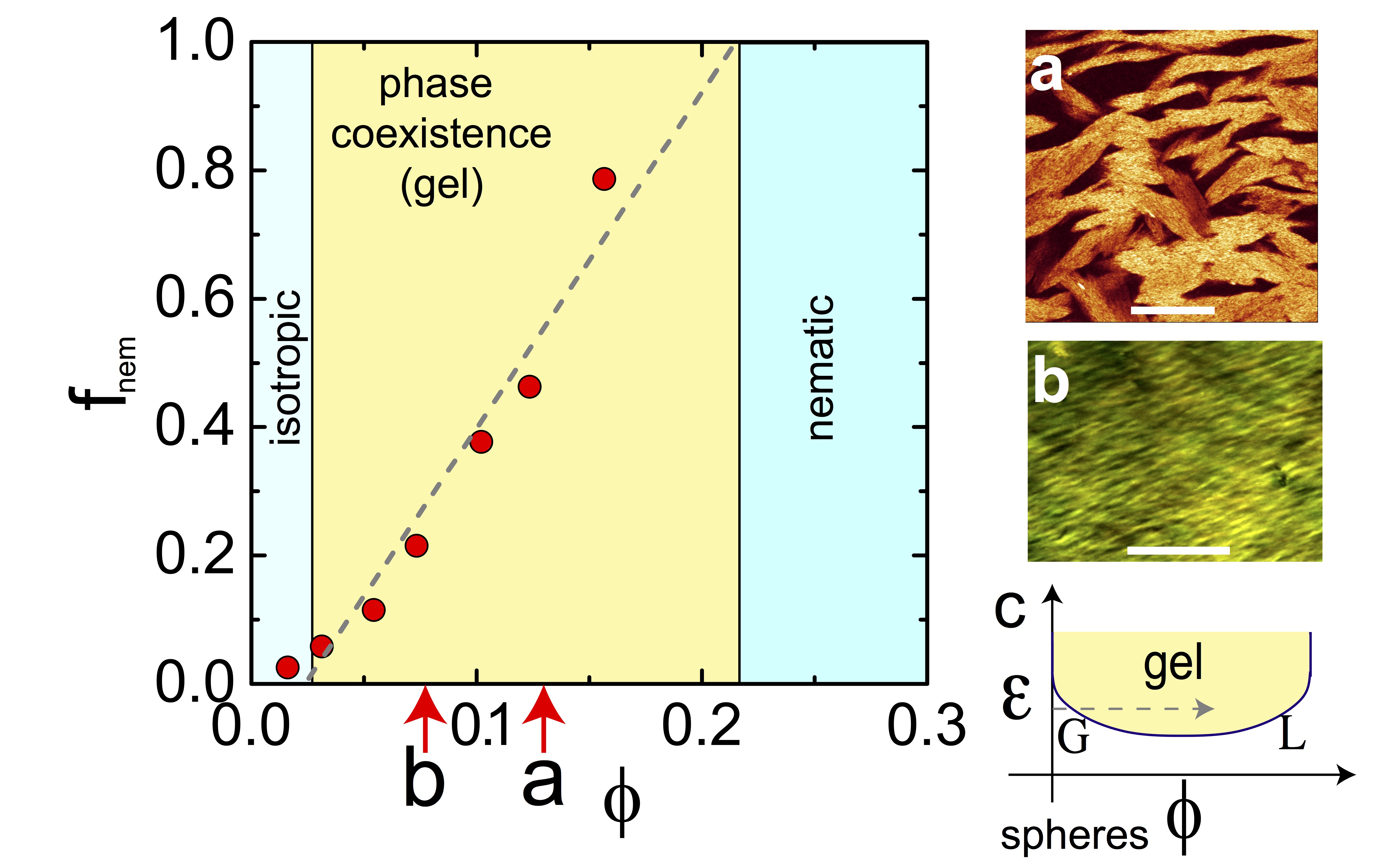}}
\caption{\label{figPhase} 
\textbf{Phase behaviour of polydisperse hard rods. }
Gelation via isotropic-nematic spinodal decomposition in polydisperse hard rods. The fraction of the nematic phase $f_\mathrm{nem}$ is plotted as a function of global volume fraction $\phi$.  Phase coexistence occurs between $\phi_\mathrm{iso}=0.024$ and $\phi_\mathrm{nem}=0.215$ for the isotropic and nematic phases respectively and it is in this region of the phase diagram that gelation occurs (yellow shaded region). 
Red data points are experimental data where we have determined $f_\mathrm{nem}$.
(a) confocal image of nematic gel at volume fraction $\phi=0.13$, bright regions of nematic phase. 
Scale bar represents $4 \; \mathrm{\mu m}$. 
(b) nematic gel imaged on a polarising optical microscope at $\phi=0.089$, a comparable state point to that shown in (a). Scale bar represents $100 \; \mathrm{\mu m}$.
(c) Gelation in a typical system of spheres. Shown is the phase diagram in the attraction strength $\varepsilon$---volume fraction $\phi$ plane.
Here gelation also occurs via spinodal decomposition, but this requires sufficient attraction for colloidal liquid (L) - gas (G) phase separation, which leads to spinodal decomposition to a (metastable) colloidal liquid phase whose volume fraction is sufficient to exhibit slow dynamics. Hard ($\varepsilon=0$) spheres do not form gels. Yellow shaded region indicates colloidal liquid-gas phase coexistence. Dashed line indicates a path for gelation.}
\end{center}
\end{figure}

\section{Results}

Our presentation of results is as follows. First we consider the phase behaviour. While the isotropic-nematic transition of long rod-shaped particles has been known since the pioneering work of Onsager \cite{onsager1949}, our results show the importance of polydispersity on broadening the coexistence gap relating to the isotropic-nematic transition, leading to a nematic phase whose density is around nine times that of the coexisting isotropic phase. Now the viscosity of the nematic phase varies markedly with respect to the mean rod orientation \cite{allen1990,vanbruggen1998}, which has profound consequences for the behaviour of the networks we obtain.

Secondly, we consider spinodal decomposition, which we demonstrate leads to a percolating network of nematic phase. We then determine the dynamical asymmetry between the isotropic and nematic phases and find the latter to be very much more viscous than the former, for flow perpendicular to the director. Finally we consider coarsening of the network. We find behaviour broadly similar to that known for spinodal gels formed of spheres \cite{lu2008,zhang2013,testard2014}, but we emphasise that the gels we obtain may exhibit flow along (but not perpendicular to) channels comprised by the nematic phase (Fig.~\ref{figRodGelSchematic}).

\subsection{Phase Behaviour}

We present the phase diagram of our system in Fig. \ref{figPhase}. Here we consider the fraction of nematic phase $f_\mathrm{nem}$ as a function of rod volume fraction $\phi$. The data we obtain from bulk observations of phase coexistence (see SI). The key point is that the isotropic-nematic phase coexistence is very substantially broadened due to polydispersity, as indicated in the yellow shaded region in Fig. \ref{figPhase}. Such broadening due to polydispersity is in quantitative agreement with theoretical predictions for hard rods \cite{speranza2002,wensink2003}. The ratio of the (effective) volume fractions of the nematic and isotropic phases for our polydisperse system is $\phi_\mathrm{nem}^\mathrm{coex}/\phi_\mathrm{iso}^\mathrm{coex}=0.215/0.024=9.00$, while that for a monodisperse system of the same aspect ratio $L/D=25$, is just $\phi_\mathrm{nem}^\mathrm{coex}/\phi_\mathrm{iso}^\mathrm{coex}=0.157/0.127=1.22$ \cite{bolhuis1997}. Thus polydispersity massively increases the density difference between the isotropic and nematic phases.

\begin{figure*}[htb]
\begin{center}
\centerline{\includegraphics[width=0.95\textwidth]{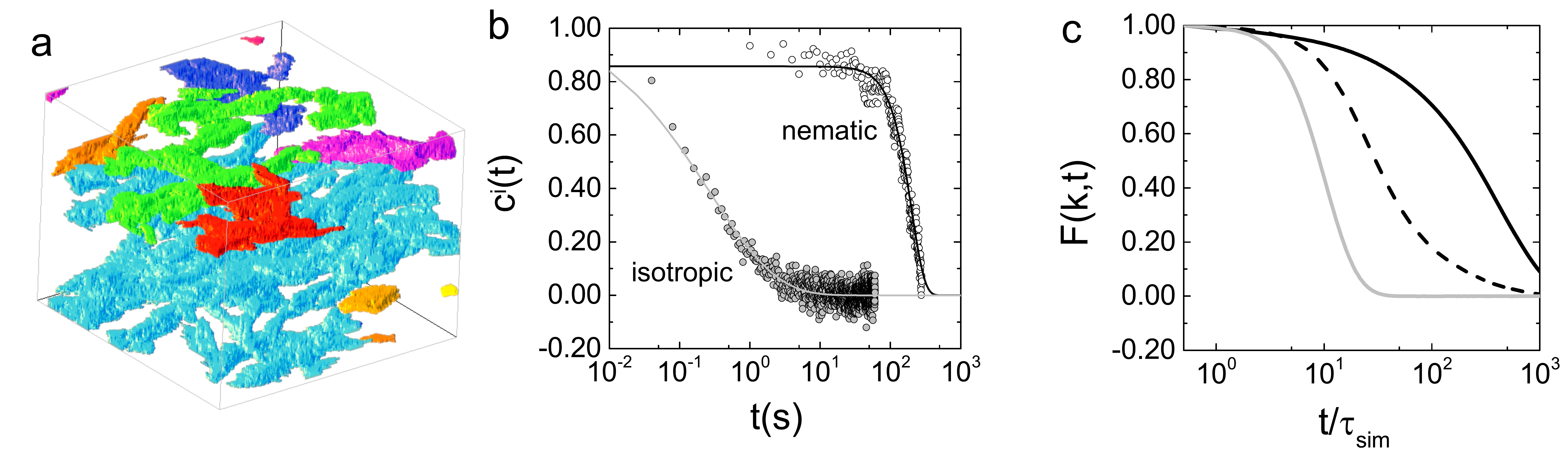}}
\caption{\label{figDynamics} 
\textbf{Evidence that polydisperse hard rods exhibit the properties required for spinodal gelation: a percolating network and dynamic asymmetry.}
(a) 3d rendering of nematic domains identified from a confocal microscopy image. Colours denote connected domains. Blue domain percolates the 3d image. $\phi=0.043$.
(b) Time-correlation functions for the isotropic and nematic phases. The correlation function $c(t)$ is detailed in the main text, and shows that the nematic phase exhibits very much slower dynamics than does the isotropic. The lines are fits to a stretched exponential from which we extract the relaxation time for each phase. 
(c) Computer simulation data for intermediate scattering functions. Here we distinguish the orientation in the case of the nematic between parallel and perpendicular to the director.}
\end{center}
\end{figure*}

A further interesting observation concerns the shape of the nematic regions as shown in the 
confocal image in Fig. \ref{figPhase}(a). Here the contrast is due to the much higher concentration of rods in the nematic compared to the isotropic phase, and the brightness levels are set such that isotropic appears dark. Nucleating nematic droplets are expected to be elongated in shape, approximately elliptical, but with sharp ends (\emph{i.e.} tactoids), as has been observed in experiments on more monodisperse systems than those we consider here \cite{oakes2007,puech2010}. Here we see a rather different geometry: the nematic regions are elongated, but there are no sharp tactoidal ends, rather a bicontinuous network with the isotropic is formed [Fig. \ref{figPhase}(a)]. We presume this is related to the spinodal decomposition that our system selects (rather than nucleation and growth) as a mechanism of phase separation, of which more below. In Fig. \ref{figPhase}(b) an image from polarising optical microscopy is shown. Here the texture of the nematic phase is rather different to that found previously, where the more monodisperse rods exhibited nucleation and growth for certain state points \cite{vanbruggen1999}.

\subsection{Spinodal decomposition}

Allied with the observation of morphology distinct from that of (isolated) tactoids anticipated in the case of nucleation and growth, we find the isotropic-nematic phase separation occurs in a spinodal-like fashion. Even at the shortest observation time accessible to our experiments  (45 s) and at the weakest supersaturation ($\phi$=0.021) we never observed nucleation and growth. That is to say, we never observed nucleation of nematic regions, these had always formed prior to our shortest observation time.

Gels often exhibit a bicontinuous texture. In Fig. \ref{figDynamics}(a) we show a 3d rendering of regions identified as nematic. We have confirmed that the regions identified as the nematic phase indeed percolate in all three dimensions 
and thus conclude that the percolation requirement for gelation is met. Close inspection of data such as that rendered in Fig. \ref{figDynamics}(a) suggests some alignment. We believe this is related to the capillary into which the sample is flowed for imaging, and may present an opportunity to produce networks whose orientation may be controlled. This novel, potentially tunable self-assembled geometry may find applications for example in microfluidic devices.

\begin{figure*}[htb]
\begin{center}
\centerline{\includegraphics[width=0.98\textwidth]{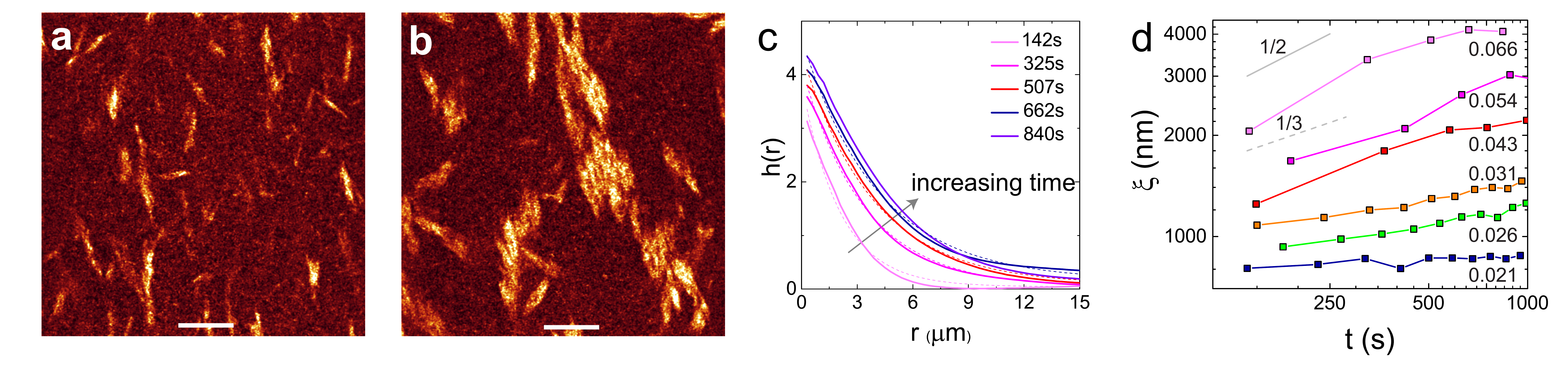}}
\caption{\label{figCoarsening} \textbf{Time-evolution of the system.} Confocal images for $\phi=0.043$, at 149 s (a) and 993 s (b) showing coarsening of the network. 
Scale bars represent $10 \; \mathrm{\mu m}$. 
(c) Pair correlation functions $h(r)$ plotted at the times shown in seconds for $\phi=0.066$ (thick solid lines). These are fitted with a decaying exponential (dashed lines) as described in the text.
(d) Quantifying coarsening with the correlation length obtained from pair correlation $h(r)$ fitting. Grey solid line indicates an exponent of $1/2$, dashed line $1/3$ as indicated.
We set the wavevector $k$ such that the lengthscale over which the dynamics are probed is comparable to that probed in the experimental data in (b).
Here the volume fraction $\phi=0.12$ and $\phi=0.36$ for the isotropic and nematic data respectively. We note that these values vary from the experiments.
}
\end{center}
\end{figure*}

\subsection{Dynamics}

We now turn to the dynamical asymmetry between the phases which is a necessary ingredient for viscoelastic phase separation, i.e. spinodal gelation \cite{tanaka2000viscoelastic}. We shall see that our system exhibits a rather unusual class of dynamic asymmetry, due to the anisotropic dynamics of the nematic phase. To measure the dynamical behaviour in our experiments, we use a time correlation function $c(t)$, which is based on pixel intensities and developed from that used for differential Fourier imaging  \cite{buzzaccaro2015}. In Fig. \ref{figDynamics}(b) we fit the time correlation function according to a stretched exponential form $c(t)=c_0 \exp[-(t/\tau)^b]$ where $c_0$ and $b$ are constants to obtain a measure of the structural relaxation time $\tau$. From our fitting, we determine relaxation times of $\tau=270$ ms for the isotropic and $\tau=202$ s for the nematic, thus indicating a considerable degree of dynamic asymmetry of three orders of magnitude. See Methods and Appendix for further details. Our experimental data suggest that this dynamical asymmetry is comparable to the differences in dynamic properties of fast and slow phases in gels made of spheres \cite{zhang2013}. However, simulation data reported in Fig. \ref{figDynamics}(c) suggest that the situation is profoundly different to gels formed of spheres: the dynamics of the nematic phase are such that the relaxation time perpendicular to the director is around a thousand times longer than along the director. This leads to a long-lived network of flowing channels. The main result here is qualitative: nematic and isotropic phases have strong dynamical contrast and thus we argue that both key ingredients for gelation are present: dynamical contrast between the two phases and spinodal decomposition leading to a bicontinuous network of nematic and isotropic.

Now the volume fractions in the simulations are rather higher than the corresponding experiments. However, determining \emph{effective} volume fractions in experiments is a notoriously challenging task \cite{royall2013myth}. We further leave the accurate determination of the equilibrium phase diagram of this system to the future, noting that the network we have formed is out of equilibrium, and that the experiments and simulations follow different dynamics.

\subsection{Coarsening}

A further key feature of gels formed by spinodal decomposition is that they coarsen over time, and this is governed by the dynamics of the more viscous phase \cite{tanaka2000viscoelastic,zhang2013,testard2014}. Our system is no exception and in Fig. \ref{figCoarsening} we present the coarsening behaviour. The confocal images in Figs. \ref{figCoarsening}(a) and (b) clearly show structural evolution of the nematic network for $\phi=0.043$ at $t=149$ and 993 s from the start of the experiment. To obtain a quantitative description of coarsening, we obtain a lengthscale from the domain size. To do this we fit $h(r)=g(r)-1$ with an exponential decay $h(r)=A \exp(r/\xi)$ where $A$ is a constant and $\xi$ is a correlation length which measures the extent of the nematic domains. Here $g(r)$ is the pixel based radial distribution function. See Appendix for details. Note that we have imposed a spherically symmetric lengthscale on a system of anisotropic particles. We have also investigated three-body correlation functions which probe the anisotropy of the system at a microscopic level, but few additional features are revealed (Appendix) and thus we restrict ourselves to the spherically symmetric analysis shown in Fig. \ref{figCoarsening}(c). The lengthscales resulting from fitting $h(r)$ are shown in Fig. \ref{figCoarsening}(d) for a range of rod volume fractions. At early times, for our deepest quench ($\phi=0.066$), the initial growth rate has an exponent $>1/2$, as indicated by the solid line in Fig. \ref{figCoarsening}(d). This is faster growth than for gels formed of spheres \cite{testard2011,zhang2013}. However at longer times and for weaker quenches, the growth rate is reduced and for certain values is compatible for that of diffusive growth of $1/3$ (dashed line in Fig. \ref{figCoarsening}(d)).

\section{Discussion}

We have argued that the properties of particulate gels, which have until now been associated with systems of \emph{attractive} particles can in fact be realised with  \emph{hard} particles in the form of polydisperse rods. By conceptual arguments, based on the dynamical contrast between the isotropic and nematic phases and spinodal demixing, we have made the case that such gels may be found in hard rods of sufficient polydispersity, with the novel feature that the viscosity of the nematic network is anisotropic: material can flow 
along the interior of the ``arms'' of the network.

We have then realised this prediction using a colloidal model system of such polydisperse hard rods of sepiolite clay, and presented four key pieces of evidence in support of our claim. Firstly, the phase coexistence in these polydisperse hard rods is broad enough that the density of the nematic phase (some 9 times higher than the coexisting isotropic phase) is sufficient that significant dynamic contrast between the phases is expected. Secondly, we have shown that the nematic phase percolates. Thirdly, we determine the dynamic contrast, with the nematic phase being very much more viscous than the isotropic phase in our experiments, but our simulation data reveals strong anisotropy in the dynamics of the nematic phase. In particular, the rods can flow along the director, but exhibit significant dynamic slowing perpendicular to the director. Finally, we have shown coarsening behaviour of the nematic domains, which is characteristic of domain coarsening in spinodal gels. We thus demonstrate a new class of ``non-sticky'' gels.

We note a novel feature of our system distinct from gels formed of spheres. In the case of spheres, as indicated in the phase diagram in the inset of Fig. \ref{figPhase}, the density of the colloid-rich phase is a strong function of attraction strength and at high density the dynamics of spheres are a strong function of density \cite{zhang2013}. Thus, (in addition to the effects of changing the interactions between the particles), the attraction strength provides a parameter by which the density of the colloid-rich network may be controlled. In particular, one finds that moving deeper in the gel region of the phase diagram upon increasing the attraction, that the rate of coarsening slows drastically, owing to the slower dynamics of the increasingly dense colloid-rich phase of spheres \cite{tanaka2000viscoelastic,zhang2013}.

In the case of our rods, the situation is profoundly different. In this athermal system, the volume fraction of the nematic phase is fixed at $\phi=0.215$. We see in Fig. \ref{figCoarsening} that, rather than slowing down upon moving deeper into the gel region by increasing $\phi$, the rate of coarsening actually \emph{accelerates}. We presume this is due either to the increased thermodynamic driving force for phase separation upon increasing $\phi$ or due to some coupling between the dynamic anisotropy of the nematic phase and the size of the domains. It would be very interesting to explore whether the same behaviour might occur in the case of spheres by moving horizontally across the gel region of the phase diagram in the $(\varepsilon,\phi)$ plane, as indicated by the dashed black line in Fig. \ref{figPhase}c, rather than vertically by changing the attraction strength $\varepsilon$ as is usually done.

It is possible to add attraction between the particles to this system by adding polymers, and this has been done for colloidal rods \cite{wilkins2009}. For small amounts of polymer, we expect a broadening of the isotropic-nematic phase coexistence \cite{tuinier2007}. Under these conditions we expect that the gelation we observe here would be even more marked because the dynamic contrast between the phases would be even larger, but qualitatively similar. However upon addition of sufficient polymer leading to stronger attraction, we expect that gels may form due to an isotropic-isotropic instability reminiscent of those which form in spheres due to a liquid-gas demixing. This would be expected to occur at rather lower volume fractions than we observe here \cite{tuinier2007}. Such attraction-driven gels of colloidal rods have indeed been observed \cite{wilkins2009}. We have also observed that no tactoids are seen, although such features are often associated with isotropic-nematic demixing of monodisperse rods \cite{oakes2007,puech2010}. Given that systems of colloidal rods in many applications are rather polydisperse, it is clearly important to establish conditions under which tactoids are actually observed. This further leads to the question as to whether a \emph{more} polydisperse system than those we have considered would in fact lead to a nematic phase with even slower dynamics than that we observe here.

A natural extension of this work is to enquire whether such behaviour is restricted to rod-like particles. We expect that this ``non-sticky'' gelation may be exhibited by a variety of anisotropic particles, which exhibit a phase coexistence gap such that the viscosity of the coexisting phases is sufficiently different. The dynamic anisotropy of course depends on the shape of the particles, but we expect that plate-like particles may exhibit comparable behaviour, if the coexistence gap between their isotropic and columnar phases is large enough. More generally the phase behaviour of a large variety of anisotropic hard particles has recently been calculated \cite{damasceno2012}. Determination of coexistence gaps, and particularly dynamic contrast between their coexisting phases, likely in the case of polydispersity, may show that a wide range of particle shapes exhibit non-sticky gelation.

\section*{Methods}

\emph{Sample preparation. --- }
Colloidal rod suspensions were made using sepiolite mineral clay particles. The zeolitic water was displaced by the fluorescent dye acridine orange \cite{yasarawan2010}. The dye-doped particles were treated with surfactant cetyltrimethylammonium bromide (CTAB, $\mathrm{C_{19}H_{42}BrN}$, BDH) solution in deionized water and centrifuged. The clay particles were dispersed in toluene and stabilized using a polymer coating of SAP230 (Infineum, UK). Further details can be found in the SI.

\emph{Confocal imaging. --- }
Confocal images of suspensions of fluorescently labeled sepiolite particles were obtained with a Leica TCS confocal microscope using a white light laser emitting at 500 nm. Borosilicate glass capillaries with cross sections of $1 \times 0.1$ mm were filled with rod suspensions at volume fractions of $0.02 < \phi < 0.07$ and glued to microscope slides with epoxy. The samples were stirred using a vortex mixer for one minute before filling the capillaries, $t=0$~s is defined as the moment when the stirring stops. The pixel size was close to 150 nm, which is around five times the rod diameter. This sets the lengthscale over which the dynamics are probed in Fig. \ref{figDynamics}(b).

\emph{Time correlation functions. --- }
The difference in dynamics between the isotropic and the nematic phase was characterised using time sequences of $xy$ images. The time resolved correlation (TRC) technique \cite{buzzaccaro2015} was used in images at $t'$ and $t' + t$ values, where $t$ represents the time over which the correlation is made. This technique measures the change in configuration by calculating the degree of correlation in the images. This correlation is calculated using individual pixel intensity values $I(\mathbf{r},t)$ of the images captured, and can be written as
\begin{equation}
C^i(t,t')=\frac{\langle I(\mathbf{r},t') I(\mathbf{r},t'+t) \rangle _\mathrm{pix} } {\langle I(\mathbf{r},t')\rangle _\mathrm{pix} \langle I(\mathbf{r},t'+t) \rangle _\mathrm{pix} } -1.
\end{equation}
$\langle \rangle _\mathrm{pix}$ indicates average over all the pixels in the image. The correlation index $C^i(t, t')$ can be normalized as $c(t, t')= C^i(t, t')/C^i(0, t')$. 
to obtain a measurement of the relaxation time in each phase.

To obtain fully demixed isotropic and nematic phases, a suspension in the coexistence regime was allowed to phase separate in a capillary for 48 h. This compares to the timescale of the gel which is several hours. Each phase was imaged far from the interface and the walls of the capillary. Two different time steps $\tau$ were chosen: 0.020 s for the isotropic phase and 1 s for the nematic phase.  The data shown in Fig. \ref{figDynamics}(b) has been corrected to account for noise in the intensity measured with the confocal microscope. This correction was made by normalizing the $c(t)$ values obtained with the first point in the correlation curve. To make the dynamic contrast clear, we subtract the constant value that $c(t\rightarrow\infty)$ approaches at long times, prior to the fitting in Fig. \ref{figDynamics}b.

\emph{Simulation model. --- } We modelled our colloidal rods as linear rigid bodies composed of several spheres that interact only with the spheres of neighbouring rods via a Weeks-Chandler-Andersen potential,
\begin{equation}
U_{WCA}(r)= \left\{
  \begin{array}{l l}
     4 \epsilon \left[ \left( \frac{\sigma}{r} \right)^{12}- \left(\frac{\sigma}{r} \right)^{6} \right]+\epsilon & \quad r < 2^{1/6} \sigma, \\
     0 & \quad r \geq 2^{1/6} \sigma,
  \end{array} \right.
\end{equation}
where $r$ is the center-of-mass distance between two spheres, $\sigma$ is the approximated diameter of the repulsive core and $\epsilon$ is the strength of the interaction in units of $k_{B}T$. For simplicity we have set $\epsilon=1$ and $\sigma=1$, and the unit of time $\tau_\mathrm{sim}=\sqrt{m\sigma^2/k_BT}$. 
Two spheres were used for each rod segment of length $1\sigma$, giving a total of 3117060 spheres for the small box of 65000 rods and 8632884 spheres for the big box of 180000 rods. The packing fractions reported were calculated by modelling the rods as hard spherocylinders with diameter $D=\sigma =1$ which gives an approximation to an effective hard body core.

For our simulations we used the open source MD simulation package LAMMPS \cite{plimpton1995lammps}, which has a dynamical integrator for rigid bodies \cite{miller2002,kamberaj2005}. To simulate our experimental conditions, we equilibrated first a system of polydisperse rods at a low packing fraction ($\phi \ll 0.01$) in a NVT ensemble using a N\'{o}se-Hoover thermostat with chains \cite{martyna1992nose}. After this, an NPT ensemble with a N\'{o}se-Hoover barostat and thermostat with chains was used to reach the desired concentrations. A final equilibration run was carried out in an NVT ensemble as in the first equilibration. Periodic boundary conditions are always applied. While here we use MD, in contrast with the dynamics of the experimental system (Brownian dynamics with hydrodynamic interactions), we have shown recently that gels of attractive spheres can reproduce experiments with MD \cite{royall2018}.

\emph{Local order parameter. --- }In the coexistence region the nematic clusters were identified by calculating the local orientational order parameter \cite{cuetos2007} of a rod $i$, which is defined by 
\begin{equation}
S_i=\frac{1}{n_i} \displaystyle\sum_{j=1}^{n_i} \left( \frac{3}{2} |\hat{\mathrm{\mathbf{e}}}_i \cdot \hat{\mathrm{\mathbf{e}}}_j| - \frac{1}{2} \right),
\end{equation}
where $\hat{\mathrm{\mathbf{e}}}_{\alpha}$ is the orientation vector of a particle $\alpha=i,j$. $n_j$ is the number of rods $j$ which are at a surface-to-surface distance of closest approach $\rho_{ij}$ from a rod $i$. Here we have set $\rho_{ij}=2.0\sigma$.

\emph{Intermediate scattering function. --- }The self part of the intermediate scattering function of our simulation systems was calculated using,
\begin{equation}
F_s(k_{\alpha \beta}, t)= \frac{1}{N} \left\langle  \displaystyle\sum _{j=1}^{N} \exp \left[ i \mathrm{\mathbf{k}_{\alpha \beta}} \cdot ( \mathrm{\mathbf{r}}_{\alpha \beta j}(t) -\mathrm{\mathbf{r}}_{\alpha \beta j}(0)) \right]  \right\rangle,
\end{equation}
were N is the number of rods. To focus on the dynamics parallel and perpendicular to the nematic director $\hat{\mathrm{\mathbf{n}}}$, the simulation boxes were rotated such that the director is aligned with the z axis ($\hat{\mathrm{\mathbf{z}}}=\hat{\mathrm{\mathbf{n}}}$). The self part of the ISF is reported as a function of the time and the radial average $k_{\alpha \beta}$ of the $\mathrm{\mathbf{k}}$ vector in the planes $xy$, $xz$ and $yz$. For our simulations we chose a $k$ value of $0.4986$ for the nematic phase and $0.4983$ for the isotropic phase.

\textbf{Acknowledgements}
We are grateful to Bob Evans and Hajime Tanaka for helpful discussions and Mike Allen and Jens Eggers for a critical reading of the manuscript.
Francesco Turci is acknowledged for help with the dynamical analysis. Richard Stenner, Jacek Wasik and Azaima Razali are thanked for help with some experiments. 
CEF acknowledges CONACyT and the Mexican Government for a student scholarship. 
CPR acknowledges the Royal Society, European Research Council (ERC Consolidator Grant NANOPRS, project number 617266) and Kyoto University SPIRITS fund. EPSRC grant code EP/ H022333/1 is acknowledged for provision of the confocal microscope used in this work. 
We thank M. Perez (Tolsa, Spain) for donating the sepiolite clay and Peter Dowding (Infineum, UK) for the SAP. TEM studies were carried out in the Chemistry Imaging Facility at UoB with equipment funded by UoB and EPSRC (EP/K035746/1 and EP/M028216/1).

\appendix{}

\section{Experimental details}

\subsection{Preparation of clay suspensions} 
\label{sec:treatment}

The sepiolite clay rod like particles were fluorescently labeled with acridine orange (AO, Sigma Aldrich) \cite{yasarawan2010}. 12 g of sepiolite were dissolved in 350 mL of deionized water and stirred for 2 h, after which the mixture was immersed in a sonic bath for 1 minute (IND 500D, Ultrawave). 1.0 g of acridine orange were dissolved in 80 mL of deionized water. The dye solution was then added to the clay slurry and left stirring overnight. The mixture was transferred into glass petri dishes and air dried. The dried clay-AO mixture was then heat treated in a vacuum oven at $120 \:^{\circ} \mathrm{C}$ for 16 h. To remove the excess dye on the surface of the clay particles, the dried clay-AO mixture was subjected to Soxhlet extraction with hot methanol for 72 h \cite{yasarawan2010}. The cleaned sepiolite-AO mixture was air dried and ground.

The dye-doped particles were then treated with surfactant. 10 g of dye-doped clay and 700 mL of deionized water were stirred for 1 h followed by the addition of 200 mL of a 0.03 M cetyltrimethylammonium bromide (CTAB, $\mathrm{C_{19}H_{42}BrN}$, BDH) solution in deionized water. The mixture was stirred for 24 h and then centrifuged at $11000$ $g$ for 1 h. The supernatant was replaced with water, redispersed and centrifuged again; this cleaning procedure was repeated two more times. The final sediment was dried in a vacuum oven at $60 \:^{\circ} \mathrm{C}$ for 16 h. The treated clay was then ground.

The clay particles were dispersed in toluene and stabilized using a polymer coating. Propylene carbonate (PC, $\mathrm{C_{4}H_{6}O_{3}}$, Acros Organics) was used to disperse the particles in toluene before any polymer was added \cite{jones1983}. A 5 wt\% suspension of treated clay in dried toluene was stirred for 2 h, then 5 wt\% of PC (relative to the clay mass) was added to the mixture and left stirring overnight. The mixture was then shear mixed for 1 minute at 25 krpm (T-18 basic, Ultra-Turrax), after which it was immersed in a sonic bath for 1 minute. Then a SAP230 (Infineum, UK) solution in toluene was added at 20 wt\% concentration at a treated clay to SAP weight ratio of 2:1 and left stirring overnight. The suspension was then centrifuged at $1000 g$ for 10 min and the sediment discarded. The remaining supernatant was centrifuged at $11000$ $g$ for 1 h. The new supernatant was then replaced with clean toluene and the sediment was redispersed, then centrifuged again at $11000$ $g$ and the supernatant replaced once more with clean toluene; this cleaning procedure was repeated twice. The mass concentration of the final suspension was found by drying a small sample. A final 4 wt\% of PC, relative to the treated clay mass, was added to the colloidal suspension. The 4 wt\% concentration of PC was kept constant for all concentrations of the suspension. After long times (>2 hrs), some sedimenation was observed. In Fig. 1a in the main text, we report the volume fraction $\phi=0.13$ corresponding to the region of the sample in which the image was taken. All other data was taken before the effects of sedimentation were observable, i.e. less than 1000 s.

\subsection{Sample characterisation}

\begin{figure*}[h]
\begin{center} 
\centerline{\includegraphics[width=120mm]{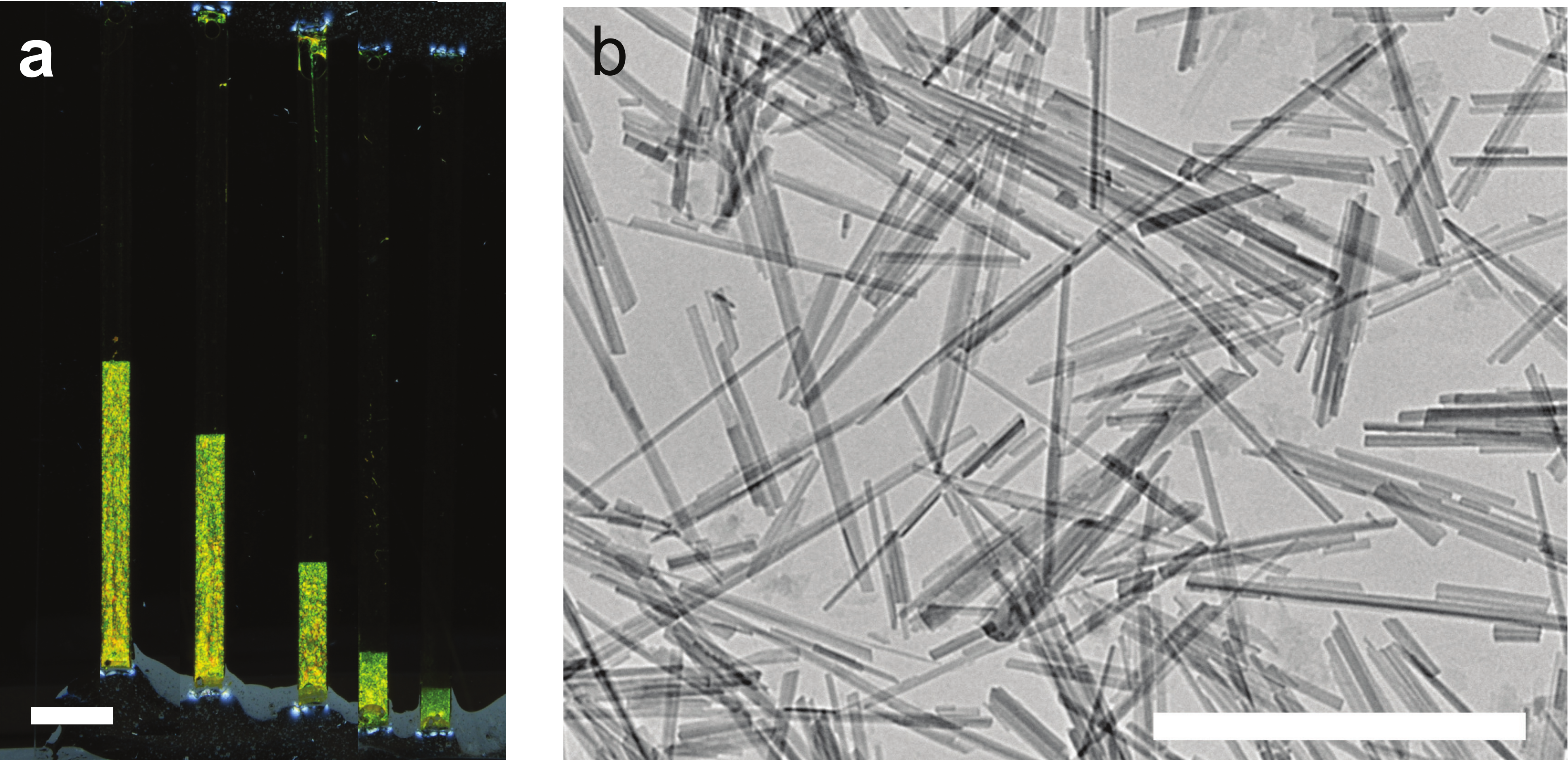}}
\end{center}
 \caption{Phase behaviour and particle characterisation. (a) Example of the images used to measure the nematic fraction, the scale bar represents 5 mm. (b) Example of TEM image used to size particles, the scale bar represents 1 $\mu$m. }
\label{sFigCharacter}
\end{figure*}

The phase diagram of the isotropic-nematic transition was obtained by filling capillaries of $50\, \times\, 2 \,\times\, 0.2$ mm with samples at different mass fractions $m$ and allowing them to phase separate for 48 h. Photographs of the capillaries between crossed polarisers were used to calculate the nematic fraction for each $m$ by measuring the height ratio of the nematic phase. Figure~\ref{sFigCharacter}a shows an example of the images used.

Transmission electron microscopy (TEM, JEOL JEM 1200-EX) was used to measure the particle dimensions (Fig. \ref{sFigCharacter}). The sample was prepared by drying a drop of the dispersion at 0.01 wt\% onto a coated copper grid. Diameter $\langle D\rangle$ and length $\langle L\rangle$ average values were obtained by measuring at least 120 particles for each quantity. The values obtained from TEM do not include the thickness of the polymer layer, which extends when the particles are in toluene. The thickness of the layer was assumed to be $\delta=4 \, \mathrm{nm}$ \cite{smits1989}, with effective particle dimensions of $\langle D' \rangle=\langle D \rangle +2\delta$ and $\langle L'\rangle =\langle L \rangle+2\delta$. The dimensions measured for the particles used in the present work were $\langle L'\rangle =723 \pm 255 \; \mathrm{nm}$ and  $\langle D'\rangle=30 \pm 6 \; \mathrm{nm}$. To account for the polydispersity, the effective aspect ratio $\langle L^{\prime}/D^{\prime} \rangle$ and its corrected standard deviation $\sigma_{LD} $ were calculated as $24.6 \pm 9.5$ using a second-order Taylor expansion;

\begin{eqnarray}
\left\langle \frac{L^{\prime}}{D^{\prime}} \right\rangle & = &\frac{ \langle L^{\prime} \rangle}{\langle D^{\prime} \rangle} \left( 1 + \left( \frac{\sigma_{D}}{\langle D^{\prime} \rangle} \right)^2 \right),\\
\sigma_{LD}&=&\frac{ \langle L^{\prime} \rangle}{\langle D^{\prime} \rangle} \sqrt{ \left( \frac{\sigma_{L}}{\langle L^{\prime} \rangle} \right)^2 + \left( \frac{\sigma_{D}}{\langle D^{\prime} \rangle} \right)^2}.
\end{eqnarray}

\noindent where $\sigma_D$ is the standard deviation of $\langle D \rangle$.

To determine the effective volume fraction $\phi$, we assumed that the particles have an effective shape of a cylinder of diameter $\langle D \rangle$ and length $\langle L \rangle$ enclosing the cuboid. The effective volume fraction is then calculated as
\begin{equation}\label{eq:voleff}
\phi= \frac{\pi}{4}\phi_{\mathrm{bare}} \left[ 1 + \frac{2 \delta}{\langle D \rangle} \left( 1+ \left( \frac{\sigma_D}{\langle D \rangle} \right) ^2 \right) \right] ^2 ,
\end{equation}
where we have taken into account polydispersity and where $\phi_{\mathrm{bare}}$ is the true volume fraction neglecting the electrostatics.

\section{Determination of two-body and three-body correlations}

The radial distribution function $g(r)$ and the three body correlation function $g_3(r,\theta)$ were calculated for pixels that represent the nematic regions. To do this, each image was binarised into nematic and isotropic according to the pixel intensity relative the average intensity for each image. Pair correlations were then determine based on pixels identified as nematic \cite{ferrierocordovathesis,royall2002}.

In the case of the three body correlation function, we chose to use equal separation $r$ and two different angles between the two sets of connected pixels $\theta=\pi, \pi/2$ \cite{royall2015}. Each 3d image had approximately $3.5 \; \times 10^8$ voxels. To accelerate the calculations, $g(r)$ and $g_3(r,\theta)$ were obtained using $1\; \times \; 10^5$ and $5\; \times \; 10^5$ white pixels respectively. The correlations obtained from those calculations are related with the structure of the nematic region and not with the orientation of the rods whose width lies well below the resolution of the confocal microscope.

\begin{figure*}[ht!]
\begin{center} 
\centerline{\includegraphics[width=100mm]{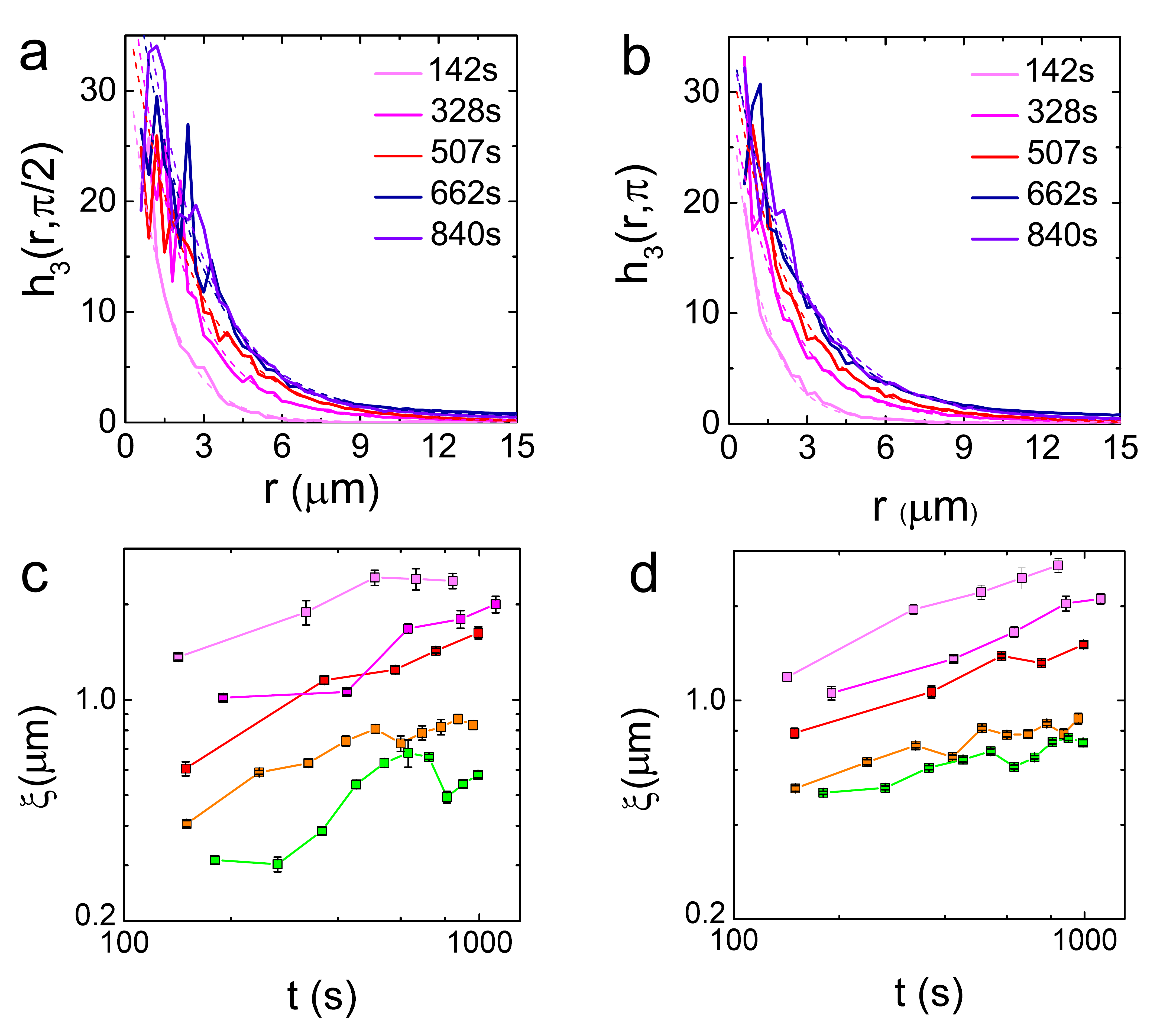}}
\end{center}
\caption{3-body correlation functions and corresponding correlation lengths. (a,b) $g_3(r,r,\theta)$ for $\phi=0.066$. (a) $g_3(r,r,\pi/2)$, (b)  $g_3(r,r,\pi)$.
Fits are decaying exponential (dashed lines).
(c,d) Growing lengthscales from fitting. (c) $g_3(r,r,\pi/2)$, (d)  $g_3(r,r,\pi)$.}
\label{sFigG3}
\end{figure*}

To obtain characteristic lengths of the nematic regions $\xi$, we fit $h(r)=g(r)-1$ and $h_3(r,\theta)=g_3(r,\theta)-1$ with a exponential decay. The results for the two-point (pair) correlation function $h(r)$ are shown in Fig. 3c of the main text. Figure \ref{sFigG3} shows equivalent results for $h_3(r,\pi/2)$ and $h_3(r,\pi)$. The correlation function with bond angle $\pi$, $h_3(r,\pi)$ does give some indication that the nematic domains are somewhat anisotropic (Fig. \ref{sFigG3}d), compared to the case wher ethe angle is $\pi/2$  (Fig. \ref{sFigG3}c) but the effect is not extreme. Moreover the values obtained are in general smaller to the ones obtained from $h(r)$.  We conclude that the nematic region exhibits some anisotropy, perhaps because their typical separation of nematic regions is comparable to or less than their length. As is the case with the two-body correlations (Fig. 3), 
we do observe an increase in the size of the nematic regions with time.

\bibliographystyle{naturemag}
\bibliography{nonsticky}

\begin{thebibliography}{10}
\expandafter\ifx\csname url\endcsname\relax
  \def\url#1{\texttt{#1}}\fi
\expandafter\ifx\csname urlprefix\endcsname\relax\def\urlprefix{URL }\fi
\providecommand{\bibinfo}[2]{#2}
\providecommand{\eprint}[2][]{\url{#2}}

\bibitem{zaccarelli2007}
\bibinfo{author}{Zaccarelli, E.}
\newblock \bibinfo{title}{Colloidal gels: Equilibrium and non-equilibrium
  routes}.
\newblock \emph{\bibinfo{journal}{J. Phys.: Condens. Matter}}
  \textbf{\bibinfo{volume}{19}}, \bibinfo{pages}{323101}
  (\bibinfo{year}{2007}).

\bibitem{tanaka2000viscoelastic}
\bibinfo{author}{Tanaka, H.}
\newblock \bibinfo{title}{Viscoelastic phase separation}.
\newblock \emph{\bibinfo{journal}{J. Phys.: Condens. Matter}}
  \textbf{\bibinfo{volume}{12}}, \bibinfo{pages}{R207} (\bibinfo{year}{2000}).

\bibitem{cardinaux2007}
\bibinfo{author}{Cardinaux, F.}, \bibinfo{author}{Gibaud, T.},
  \bibinfo{author}{Stradner, A.} \& \bibinfo{author}{Schurtenberger, P.}
\newblock \bibinfo{title}{Interplay between spinodal decomposition and glass
  formation in proteins exhibiting short-range attractions}.
\newblock \emph{\bibinfo{journal}{Phys. Rev. Lett.}}
  \textbf{\bibinfo{volume}{99}}, \bibinfo{pages}{118301}
  (\bibinfo{year}{2007}).

\bibitem{jabbarifarouji2007}
\bibinfo{author}{Jabbari-Farouji, S.}, \bibinfo{author}{Wegdam, G.} \&
  \bibinfo{author}{Bonn, D.}
\newblock \bibinfo{title}{Gels and glasses in a single system: Evidence for an
  intricate free-energy landscape of glassy materials}.
\newblock \emph{\bibinfo{journal}{Phys. Rev. Lett.}}
  \textbf{\bibinfo{volume}{99}}, \bibinfo{pages}{065701}
  (\bibinfo{year}{2007}).

\bibitem{tanaka2013fara}
\bibinfo{author}{Tanaka, H.}
\newblock \bibinfo{title}{Viscoelastic phase separation in soft matter and
  foods}.
\newblock \emph{\bibinfo{journal}{Faraday Discuss.}}
  \textbf{\bibinfo{volume}{167}}, \bibinfo{pages}{9--76}
  (\bibinfo{year}{2013}).

\bibitem{helgeson2012}
\bibinfo{author}{Helgeson, M.~E.}, \bibinfo{author}{Moran, S.~E.},
  \bibinfo{author}{An, H.~Z.},  \& \bibinfo{author}{Doyle, P.~S.}
\newblock \bibinfo{title}{Mesoporous organohydrogels from thermogelling
  photocrosslinkable nanoemulsions}.
\newblock \emph{\bibinfo{journal}{Nature Mater.}}
  \textbf{\bibinfo{volume}{11}}, \bibinfo{pages}{344--352}
  (\bibinfo{year}{2012}).

\bibitem{rose2014}
\bibinfo{author}{Rose, S.} \emph{et~al.}
\newblock \bibinfo{title}{Nanoparticle solutions as adhesives for gels and
  biological tissues}.
\newblock \emph{\bibinfo{journal}{Nature}} \textbf{\bibinfo{volume}{505}},
  \bibinfo{pages}{382--385} (\bibinfo{year}{2014}).

\bibitem{ulrich2009}
\bibinfo{author}{Ulrich, S.} \emph{et~al.}
\newblock \bibinfo{title}{Cooling and aggregation in wet granulates}.
\newblock \emph{\bibinfo{journal}{Phys. Rev. Lett.}}
  \textbf{\bibinfo{volume}{102}}, \bibinfo{pages}{148002}
  (\bibinfo{year}{2009}).

\bibitem{bouttes2014}
\bibinfo{author}{Bouttes, D.}, \bibinfo{author}{Gouillart, E.},
  \bibinfo{author}{Boller, E.}, \bibinfo{author}{Dalmas, D.} \&
  \bibinfo{author}{Vandembroucq, D.}
\newblock \bibinfo{title}{Fragmentation and limits to dynamical scaling in
  viscous coarsening: An interrupted in situ x-ray tomographic study}.
\newblock \emph{\bibinfo{journal}{Phys. Rev. Lett.}}
  \textbf{\bibinfo{volume}{112}}, \bibinfo{pages}{245701}
  (\bibinfo{year}{2014}).

\bibitem{baumer2013}
\bibinfo{author}{Baumer, R.~E.} \& \bibinfo{author}{Demkowicz, M.~J.}
\newblock \bibinfo{title}{Glass transition by gelation in a phase separating
  binary alloy}.
\newblock \emph{\bibinfo{journal}{Phys. Rev. Lett.}}
  \textbf{\bibinfo{volume}{110}}, \bibinfo{pages}{145502}
  (\bibinfo{year}{2013}).

\bibitem{hsiao2012}
\bibinfo{author}{Hsiao, L.~C.}, \bibinfo{author}{Newman, R.~S.},
  \bibinfo{author}{Glotzer, S.~C.} \& \bibinfo{author}{Solomon, M.~J.}
\newblock \bibinfo{title}{Role of isostaticity and load-bearing microstructure
  in the elasticity of yielded colloidal gels}.
\newblock \emph{\bibinfo{journal}{Proc. Nat. Acad. Sci.}}
  \textbf{\bibinfo{volume}{109}}, \bibinfo{pages}{16029--16034}
  (\bibinfo{year}{2012}).

\bibitem{zhang2012}
\bibinfo{author}{Zhang, T.~H.}, \bibinfo{author}{Klok, J.},
  \bibinfo{author}{Tromp, R.~H.}, \bibinfo{author}{Groenewold, J.} \&
  \bibinfo{author}{Kegel, W.~K.}
\newblock \bibinfo{title}{Non-equilibrium cluster states in colloids with
  competing interactions}.
\newblock \emph{\bibinfo{journal}{Soft Matter}} \textbf{\bibinfo{volume}{8}},
  \bibinfo{pages}{667} (\bibinfo{year}{2012}).

\bibitem{valadezperez2013}
\bibinfo{author}{Valadez-P\'{e}rez, N.~E.}, \bibinfo{author}{Liu, Y.},
  \bibinfo{author}{Eberle, A. P.~R.}, \bibinfo{author}{Wagner, N.~J.} \&
  \bibinfo{author}{Casta\~{n}da Priego, R.}
\newblock \bibinfo{title}{Dynamical arrest in adhesive hard-sphere dispersions
  driven by rigidity percolation}.
\newblock \emph{\bibinfo{journal}{Phys. Rev. E}} \textbf{\bibinfo{volume}{88}},
  \bibinfo{pages}{060302(R)} (\bibinfo{year}{2013}).

\bibitem{varrato2012}
\bibinfo{author}{Varrato, F.} \emph{et~al.}
\newblock \bibinfo{title}{Arrested demixing opens route to bigels}.
\newblock \emph{\bibinfo{journal}{Proc. Nat. Acad. Sci.}}
  \textbf{\bibinfo{volume}{109}}, \bibinfo{pages}{19155--19160}
  (\bibinfo{year}{2012}).

\bibitem{poon2002}
\bibinfo{author}{Poon, W. C.~K.}
\newblock \bibinfo{title}{The physics of a model colloid-polymer mixture}.
\newblock \emph{\bibinfo{journal}{J. Phys: Condens. Matter}}
  \textbf{\bibinfo{volume}{14}}, \bibinfo{pages}{R859--R880}
  (\bibinfo{year}{2002}).

\bibitem{lu2008}
\bibinfo{author}{Lu, P.~J.} \emph{et~al.}
\newblock \bibinfo{title}{Gelation of particles with short-range attraction}.
\newblock \emph{\bibinfo{journal}{Nature}} \textbf{\bibinfo{volume}{453}},
  \bibinfo{pages}{499--504} (\bibinfo{year}{2008}).

\bibitem{royall2018}
\bibinfo{author}{Royall, C.~P.}, \bibinfo{author}{Williams, S.~R.} \&
  \bibinfo{author}{Tanaka, H.}
\newblock \bibinfo{title}{Vitrification and gelation in sticky spheres}.
\newblock \emph{\bibinfo{journal}{J. Chem. Phys.}}
  \textbf{\bibinfo{volume}{148}}, \bibinfo{pages}{044501}
  (\bibinfo{year}{2018}).
\newblock
  \urlprefix\url{http://scitation.aip.org/content/aip/journal/jcp/148/4/10.1063/1.5000263}.

\bibitem{zhang2013}
\bibinfo{author}{Zhang, I.}, \bibinfo{author}{Royall, C.~P.},
  \bibinfo{author}{Faers, M.~A.} \& \bibinfo{author}{Bartlett, P.}
\newblock \bibinfo{title}{Phase separation dynamics in colloid-polymer
  mixtures: the effect of interaction range}.
\newblock \emph{\bibinfo{journal}{Soft Matter}} \textbf{\bibinfo{volume}{9}},
  \bibinfo{pages}{2076--2084} (\bibinfo{year}{2013}).

\bibitem{vanbruggen1999}
\bibinfo{author}{van Bruggen, M. P.~B.}, \bibinfo{author}{Dhont, J. K.~G.} \&
  \bibinfo{author}{Lekkerkerker, H. N.~W.}
\newblock \bibinfo{title}{Morphology and kinetics of the isotropic-nematic
  phase transition in dispersions of hard rods}.
\newblock \emph{\bibinfo{journal}{Macromolecules}}
  \textbf{\bibinfo{volume}{32}}, \bibinfo{pages}{2256--2264}
  (\bibinfo{year}{1999}).

\bibitem{ni2010}
\bibinfo{author}{Ni, R.}, \bibinfo{author}{Belli, S.}, \bibinfo{author}{van
  Roij, R.} \& \bibinfo{author}{Dijkstra, M.}
\newblock \bibinfo{title}{Glassy dynamics, spinodal fluctuations, and the
  kinetic limit of nucleation in suspensions of colloidal hard rods}.
\newblock \emph{\bibinfo{journal}{Phys. Rev. Lett.}}
  \textbf{\bibinfo{volume}{105}}, \bibinfo{pages}{088302}
  (\bibinfo{year}{2010}).

\bibitem{lettinga2006}
\bibinfo{author}{Lettinga, M.~P.} \emph{et~al.}
\newblock \bibinfo{title}{Nematic-isotropic spinodal decomposition kinetics of
  rodlike viruses}.
\newblock \emph{\bibinfo{journal}{Phys. Rev. E}} \textbf{\bibinfo{volume}{73}},
  \bibinfo{pages}{011412} (\bibinfo{year}{2006}).

\bibitem{woolston2015}
\bibinfo{author}{Woolston, P.} \& \bibinfo{author}{van Duijneveldt, J.~S.}
\newblock \bibinfo{title}{Isotropic-nematic phase transition of polydisperse
  clay rods}.
\newblock \emph{\bibinfo{journal}{J. Chem. Phys.}}
  \textbf{\bibinfo{volume}{142}}, \bibinfo{pages}{184901}
  (\bibinfo{year}{2015}).

\bibitem{speranza2002}
\bibinfo{author}{Speranza, A.} \& \bibinfo{author}{Sollich, P.}
\newblock \bibinfo{title}{Simplified onsager theory for isotropic-nematic phase
  equilibria of length polydisperse hard rods}.
\newblock \emph{\bibinfo{journal}{J. Chem. Phys.}}
  \textbf{\bibinfo{volume}{117}}, \bibinfo{pages}{5421--5436}
  (\bibinfo{year}{2002}).

\bibitem{onsager1949}
\bibinfo{author}{Onsager, L.}
\newblock \bibinfo{title}{The effects of shape on the interaction of colloidal
  particles}.
\newblock \emph{\bibinfo{journal}{Ann. N. Y. Acad. Sci.}}
  \textbf{\bibinfo{volume}{51}}, \bibinfo{pages}{627--659}
  (\bibinfo{year}{1949}).

\bibitem{allen1990}
\bibinfo{author}{Allen, M.~P.}
\newblock \bibinfo{title}{Diffusion coefficient increases with density in hard
  ellipsoid liquid crystals}.
\newblock \emph{\bibinfo{journal}{Phys. Rev. Lett.}}
  \textbf{\bibinfo{volume}{65}}, \bibinfo{pages}{2881--2884}
  (\bibinfo{year}{1990}).

\bibitem{vanbruggen1998}
\bibinfo{author}{van Bruggen, M. P.~B.}, \bibinfo{author}{Lekkerkerker, H.
  N.~W.}, \bibinfo{author}{Maret, G.} \& \bibinfo{author}{Dhont, J. K.~G.}
\newblock \bibinfo{title}{Long-time translational self-diffusion in isotropic
  and nematic dispersions of colloidal rods}.
\newblock \emph{\bibinfo{journal}{Phys. Rev. E}} \textbf{\bibinfo{volume}{58}},
  \bibinfo{pages}{7668--7677} (\bibinfo{year}{1998}).

\bibitem{testard2014}
\bibinfo{author}{Testard, V.}, \bibinfo{author}{Bethier, L.} \&
  \bibinfo{author}{Kob, W.}
\newblock \bibinfo{title}{Intermittent dynamics and logarithmic domain growth
  during the spinodal decomposition of a glass-forming liquid}.
\newblock \emph{\bibinfo{journal}{J. Chem. Phys.}}
  \textbf{\bibinfo{volume}{140}}, \bibinfo{pages}{164502}
  (\bibinfo{year}{2014}).

\bibitem{wensink2003}
\bibinfo{author}{Wensink, H.~H.} \& \bibinfo{author}{Vroege, G.~J.}
\newblock \bibinfo{title}{Isotropic-nematic phase behavior of
  length-polydisperse hard rods}.
\newblock \emph{\bibinfo{journal}{J. Chem. Phys.}}
  \textbf{\bibinfo{volume}{119}}, \bibinfo{pages}{6868--6882}
  (\bibinfo{year}{2003}).

\bibitem{bolhuis1997}
\bibinfo{author}{Bolhuis, P.} \& \bibinfo{author}{Frenkel, D.}
\newblock \bibinfo{title}{Tracing the phase boundaries of hard
  spherocylinders}.
\newblock \emph{\bibinfo{journal}{J. Chem. Phys.}}
  \textbf{\bibinfo{volume}{106}}, \bibinfo{pages}{666--687}
  (\bibinfo{year}{1997}).

\bibitem{oakes2007}
\bibinfo{author}{Oakes, P.~W.}, \bibinfo{author}{Viamontes, J.} \&
  \bibinfo{author}{Tang, J.~X.}
\newblock \bibinfo{title}{Growth of tactoidal droplets during the first-order
  isotropic to nematic phase transition of f-actin}.
\newblock \emph{\bibinfo{journal}{Phys. Rev. E}} \textbf{\bibinfo{volume}{75}},
  \bibinfo{pages}{061902} (\bibinfo{year}{2007}).

\bibitem{puech2010}
\bibinfo{author}{Puech, N.}, \bibinfo{author}{Grelet, E.},
  \bibinfo{author}{Poulin, P.}, \bibinfo{author}{Blanc, C.} \&
  \bibinfo{author}{van~der Schoot, P.}
\newblock \bibinfo{title}{Nematic droplets in aqueous dispersions of carbon
  nanotubes}.
\newblock \emph{\bibinfo{journal}{Phys. Rev. E}} \textbf{\bibinfo{volume}{82}},
  \bibinfo{pages}{020702} (\bibinfo{year}{2010}).

\bibitem{buzzaccaro2015}
\bibinfo{author}{Buzzaccaro, S.}, \bibinfo{author}{Alaimo, M.~D.},
  \bibinfo{author}{Secchi, E.} \& \bibinfo{author}{Piazza, R.}
\newblock \bibinfo{title}{Spatially: resolved heterogeneous dynamics in a
  strong colloidal gel}.
\newblock \emph{\bibinfo{journal}{J. Phys.: Condens. Matter}}
  \textbf{\bibinfo{volume}{27}}, \bibinfo{pages}{194120}
  (\bibinfo{year}{2015}).

\bibitem{royall2013myth}
\bibinfo{author}{Royall, C.~P.}, \bibinfo{author}{Poon, W. C.~K.} \&
  \bibinfo{author}{Weeks, E.~R.}
\newblock \bibinfo{title}{In search of colloidal hard spheres}.
\newblock \emph{\bibinfo{journal}{Soft Matter}} \textbf{\bibinfo{volume}{9}},
  \bibinfo{pages}{17--27} (\bibinfo{year}{2013}).

\bibitem{testard2011}
\bibinfo{author}{Testard, V.}, \bibinfo{author}{Berthier, L.} \&
  \bibinfo{author}{Kob, W.}
\newblock \bibinfo{title}{Influence of the glass transition on the liquid-gas
  spinodal decomposition}.
\newblock \emph{\bibinfo{journal}{Phys. Rev. Lett.}}
  \textbf{\bibinfo{volume}{106}}, \bibinfo{pages}{125702}
  (\bibinfo{year}{2011}).

\bibitem{wilkins2009}
\bibinfo{author}{Wilkins, G. M.~H.}, \bibinfo{author}{Spicer, P.~T.} \&
  \bibinfo{author}{Solomon, M.~J.}
\newblock \bibinfo{title}{Colloidal system to explore structural and dynamical
  transitions in rod networks, gels, and glasses}.
\newblock \emph{\bibinfo{journal}{Langmuir}} \textbf{\bibinfo{volume}{25}},
  \bibinfo{pages}{8951--8959} (\bibinfo{year}{2009}).

\bibitem{tuinier2007}
\bibinfo{author}{Tuinier, R.}, \bibinfo{author}{Taniguchi, T.} \&
  \bibinfo{author}{Wensink, H.}
\newblock \bibinfo{title}{Phase behavior of a suspension of hard
  spherocylinders plus ideal polymer chains}.
\newblock \emph{\bibinfo{journal}{Eur. Phys. J. E.}}
  \textbf{\bibinfo{volume}{23}}, \bibinfo{pages}{355--365}
  (\bibinfo{year}{2007}).

\bibitem{damasceno2012}
\bibinfo{author}{Damasceno, P.~F.}, \bibinfo{author}{Engel, M.} \&
  \bibinfo{author}{Glotzer, S.~C.}
\newblock \bibinfo{title}{Predictive self-assembly of polyhedra into complex
  structures}.
\newblock \emph{\bibinfo{journal}{Science}} \textbf{\bibinfo{volume}{337}},
  \bibinfo{pages}{453--457} (\bibinfo{year}{2012}).

\bibitem{yasarawan2010}
\bibinfo{author}{Yasarawan, N.} \& \bibinfo{author}{van Duijneveldt, J.~S.}
\newblock \bibinfo{title}{Arrested phase separation of colloidal rodÐsphere
  mixtures}.
\newblock \emph{\bibinfo{journal}{Soft Matter}} \textbf{\bibinfo{volume}{6}},
  \bibinfo{pages}{353--362} (\bibinfo{year}{2010}).

\bibitem{plimpton1995lammps}
\bibinfo{author}{Plimpton, S.}
\newblock \bibinfo{title}{Fast parallel algorithms for short-range molecular
  dynamics}.
\newblock \emph{\bibinfo{journal}{Journal of computational physics}}
  \textbf{\bibinfo{volume}{117}}, \bibinfo{pages}{1--19}
  (\bibinfo{year}{1995}).

\bibitem{miller2002}
\bibinfo{author}{Miller~Iii, T.} \emph{et~al.}
\newblock \bibinfo{title}{Symplectic quaternion scheme for biophysical
  molecular dynamics}.
\newblock \emph{\bibinfo{journal}{The Journal of chemical physics}}
  \textbf{\bibinfo{volume}{116}}, \bibinfo{pages}{8649--8659}
  (\bibinfo{year}{2002}).

\bibitem{kamberaj2005}
\bibinfo{author}{Kamberaj, H.}, \bibinfo{author}{Low, R.} \&
  \bibinfo{author}{Neal, M.}
\newblock \bibinfo{title}{Time reversible and symplectic integrators for
  molecular dynamics simulations of rigid molecules}.
\newblock \emph{\bibinfo{journal}{The Journal of chemical physics}}
  \textbf{\bibinfo{volume}{122}}, \bibinfo{pages}{224114}
  (\bibinfo{year}{2005}).

\bibitem{martyna1992nose}
\bibinfo{author}{Martyna, G.~J.}, \bibinfo{author}{Klein, M.~L.} \&
  \bibinfo{author}{Tuckerman, M.}
\newblock \bibinfo{title}{Nos{\'e}--hoover chains: the canonical ensemble via
  continuous dynamics}.
\newblock \emph{\bibinfo{journal}{The Journal of chemical physics}}
  \textbf{\bibinfo{volume}{97}}, \bibinfo{pages}{2635--2643}.

\bibitem{cuetos2007}
\bibinfo{author}{Cuetos, A.} \& \bibinfo{author}{Dijkstra, M.}
\newblock \bibinfo{title}{Kinetic pathways for the isotropic-nematic phase
  transition in a system of colloidal hard rods: a simulation study}.
\newblock \emph{\bibinfo{journal}{Physical review letters}}
  \textbf{\bibinfo{volume}{98}}, \bibinfo{pages}{095701}
  (\bibinfo{year}{2007}).

\bibitem{jones1983}
\bibinfo{author}{Jones, T.}
\newblock \emph{\bibinfo{journal}{Clay Miner.}} \textbf{\bibinfo{volume}{18}},
  \bibinfo{pages}{399--401} (\bibinfo{year}{1983}).

\bibitem{smits1989}
\bibinfo{author}{Smits, C.}, \bibinfo{author}{Briels, W.},
  \bibinfo{author}{Dhont, J.} \& \bibinfo{author}{Lekkerkerker, H.}
\newblock \bibinfo{title}{Influence of the stabilizing coating on the rate of
  crystallization of colloidal systems}.
\newblock \emph{\bibinfo{journal}{Prog. Colloid Polym. Sci.}}
  \textbf{\bibinfo{volume}{79}}, \bibinfo{pages}{287--292}
  (\bibinfo{year}{1989}).

\bibitem{ferrierocordovathesis}
\bibinfo{author}{Ferriero~Cordova, C.~E.}
\newblock \emph{\bibinfo{title}{Stucture formation in colloidal rod
  suspensions: experiments and computer simulations}}.
\newblock Ph.D. thesis, \bibinfo{school}{University of Bristol}
  (\bibinfo{year}{2016}).

\bibitem{royall2002}
\bibinfo{author}{Royall, C.~P.} \& \bibinfo{author}{Donald, A.~M.}
\newblock \bibinfo{title}{Structure of silica in matt water-based lacquer}.
\newblock \emph{\bibinfo{journal}{Phys. Rev. E}} \textbf{\bibinfo{volume}{66}},
  \bibinfo{pages}{021406} (\bibinfo{year}{2002}).

\bibitem{royall2015}
\bibinfo{author}{Royall, C.~P.}, \bibinfo{author}{Eggers, J.},
  \bibinfo{author}{Furukawa, A.} \& \bibinfo{author}{Tanaka, H.}
\newblock \bibinfo{title}{Probing colloidal gels at multiple length scales: The
  role of hydrodynamics}.
\newblock \emph{\bibinfo{journal}{Phys. Rev. Lett.}}
  \textbf{\bibinfo{volume}{114}}, \bibinfo{pages}{258302}
  (\bibinfo{year}{2015}).

\end{thebibliography}


\end{document}